\documentstyle[12pt,psfig]{ar}
\onecolumn

\begin{document}

\thispagestyle{empty}

\title{Detection of heavy-metal lines in the spectrum of the circumstellar
       envelope of a post-AGB star}

\author{V.\,G.~Klochkova}

\date{}{}

\institute{Special Astrophysical Observatory, 
          Nizhnij Arkhyz, 369167, Russia}

\abstract{Splitting of the strongest absorption lines with a
lower-level excitation potential $\chi_{\rm low} < 1$\,eV has been detected for the
first time in the optical spectra of the post-AGB star V354\,Lac
obtained with a spectral resolution R\,=\,60000 at the 6-m telescope BTA.
Analysis of the kinematics shows that the short-wavelength component of
the split line originates in the star's thick gas--dust envelope.
Disregarding the splitting of strong lines when the chemical composition
is calculated leads to overestimated excess of s-process elements
(Ba, La, Ce, Nd) in the stellar atmosphere. The profiles of strong
absorption lines have been found to be variable. The available
radial-velocity data suggest the absence of any trend in the velocity
field in the atmosphere and circumstellar envelope of V354\,Lac over 15
years of its observations.}

\titlerunning{\it The spectrum  of a post-AGB star V354\,Lac}
		      
\authorrunning{\it V.G.\,Klochkova}
\maketitle


\section{Introduction}

The cool variable star V354\,Lac\,=\,HD\,235858 identified with the
infrared source IRAS\,22272+5435 is one of the most interesting candidates
for protoplanetary ne\--bulae (PPN). Intermediate-mass stars that evolve from
the asymptotic giant branch (AGB) to a planetary nebula are observed at
the short PPN phase of evolution. The initial masses of these stars lie
within the range 3--8${\mathcal M}_{\odot}$. The evolution of
intermediate-mass stars was described in detail, for example, by Blocker
(2001), while we will recall only the main points of this process. Having
passed through the successive core hydrogen and helium burning phases of
evolution, these stars underwent great mass loss in the form of a strong
stellar wind at the AGB phase (the mass loss rate reached
10$^{-4}{\mathcal M}_{\odot}/$yr). Since the bulk of the stellar mass is
lost, a post-AGB star is a degenerate carbon--oxygen core with a typical
mass of about 0.6${\mathcal M}_{\odot}$ surrounded by an expanding
gas--dust envelope. The interest of astronomers in PPN stems, first,
from the possibility of studying the history of mass loss via a stellar
wind and, second, from the unique opportunity to observe the result of
stellar nucleosynthesis, mixing, and dredge-up of nuclear-reaction
products to the surface layers during the preceding evolution of the star.

V354\,Lac was one of the first PPN candidates with feature at 21\,$\mu$ in
the infrared spectrum whose atmospheres exhibited large overabundances of
carbon and s--process elements (Zacs {\it et~al.}~1995). The energy
distribution of V354\,Lac has a double-peak pattern typical of PPN, with
the total energies emitted by the star in the visible wavelength range and
by the circumstellar envelope in the infrared being almost identical (see
Fig.\,4 in Hrivnak \&~Kwok~(1991)). In the group of related objects,
V354\,Lac stands out by significant photometric variability. According to
Hrivnak and Kwok~(1991), the B and V magnitudes for two epochs of
observations differed by 0$\lefteqn{.}^{\rm m}$72 and 0$\lefteqn{.}^{\rm
m}$84, respectively.

Secular variability of the main parameters detected in several PPN
stimulates a spectroscopic monitoring of the most probable PPN candidates.
For example, we detected spectroscopic variability of the optical
counterparts of the sources IRAS\,01005+7910 (Klochkova {\it et~al.} 2002a),
IRAS\,05040+4820 (Klochkova {\it et~al.} 2004a), and IRAS\,20572+4919
(Klochkova {\it et~al.} 2008) and found a trend in the effective temperature T$_{\rm eff}$
for the star HD\,161796\,=\,IRAS\,17436+5003 (Klochkova {\it et~al.} 2002b).
Here, it is also pertinent to recall the evolution of the parameters and
chemical composition of the famous highly evolved star FG\,Sge observed
for more than a century (see the review by Jeffery and Schonberner (2006)
and references therein). In this paper, we present the results of high
spectral resolution observations of V354\,Lac for the epoch 2007--2008 and
compare the new data with previous ones. Our main goal is to reveal
probable spectroscopic variability and peculiarity of the spectrum as well
as to study the velocity field in the atmosphere and circumstellar
envelope of the star.

\section{Observations and spectroscopic data reduction}

We obtained new spectroscopic data for V354\,Lac with the NES echelle
spectrograph (Panchuk {\it et~al.} 2007) at the Nasmyth focus of the 6-m
telescope BTA of the Special Astrophysical Observatory. The observations
were performed with a large-size 2048$\times$2048-pixel CCD array and an
image slicer (Panchuk {\it et~al.} 2007). The spectral resolution was
R\,=\,60000. The first spectrum (JD\,=\,2454170.58) was taken in the
wavelength range 4514--5940\,\AA{}, the next two spectra
(JD\,=\,2454225.51 and 2454727.35) were taken in the longer wavelength
range, 5215--6690 and 5260--6760 A, respectively. One-dimensional spectra
were extracted from two-dimensional echelle frames using the ECHELLE
context of the MIDAS software package modified by Yushkin \& Klochkova
(2005). Cosmic-ray particle hits were removed by a median averaging of two
successive spectra. The wavelength calibration was performed using the
spectra of a hollow-cathode Th--Ar lamp. The heliocentric radial
velocities V$_{\odot}$ estimated from these spectra and listed below in
the table were found using the DECH\,20 package (Galazutdinov 1992).

\section{Discussion of results}

\subsection{Peculiarity of the Spectrum}

\begin{figure}[h]
\psfig{figure=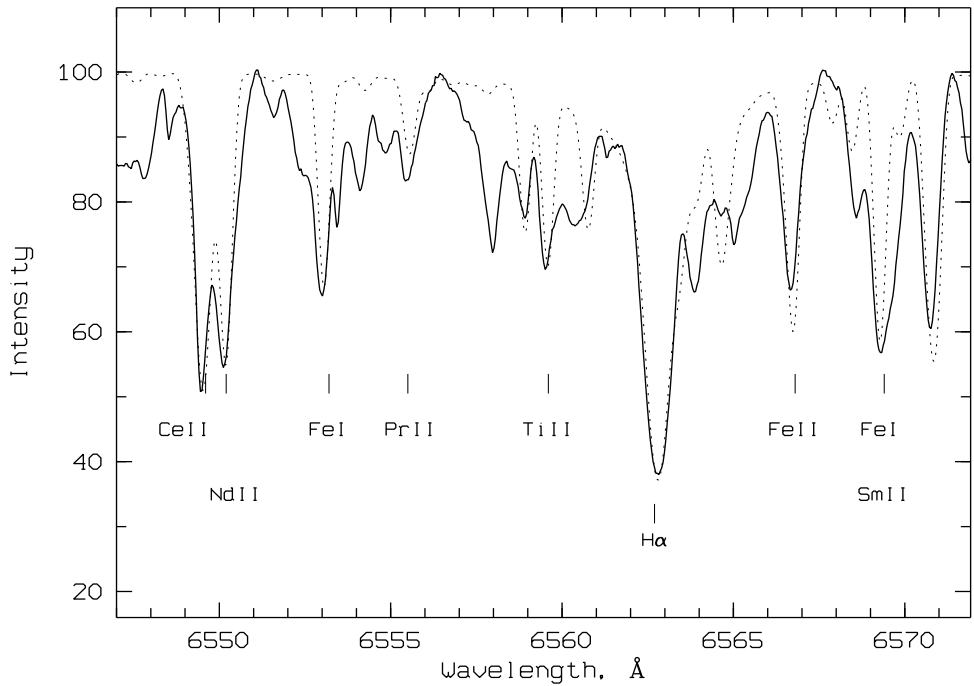,angle=-90,width=15cm,height=10cm}
\caption{Part of the spectrum for V354\,Lac near H$\alpha$. The dotted line
        indicates the theore\--tical spectrum calculated with T$_{\rm eff}$\,=\,5650\,K,
	$\log g=0.2$, $\xi_{\rm t} = 5.0$\,km$/s$, and the elemental abundances derived
        previously (Klochkova {\it et~al.} 2009). The telluric spectrum was not subtracted.}
\end{figure}

\begin{figure}[h]
\psfig{figure=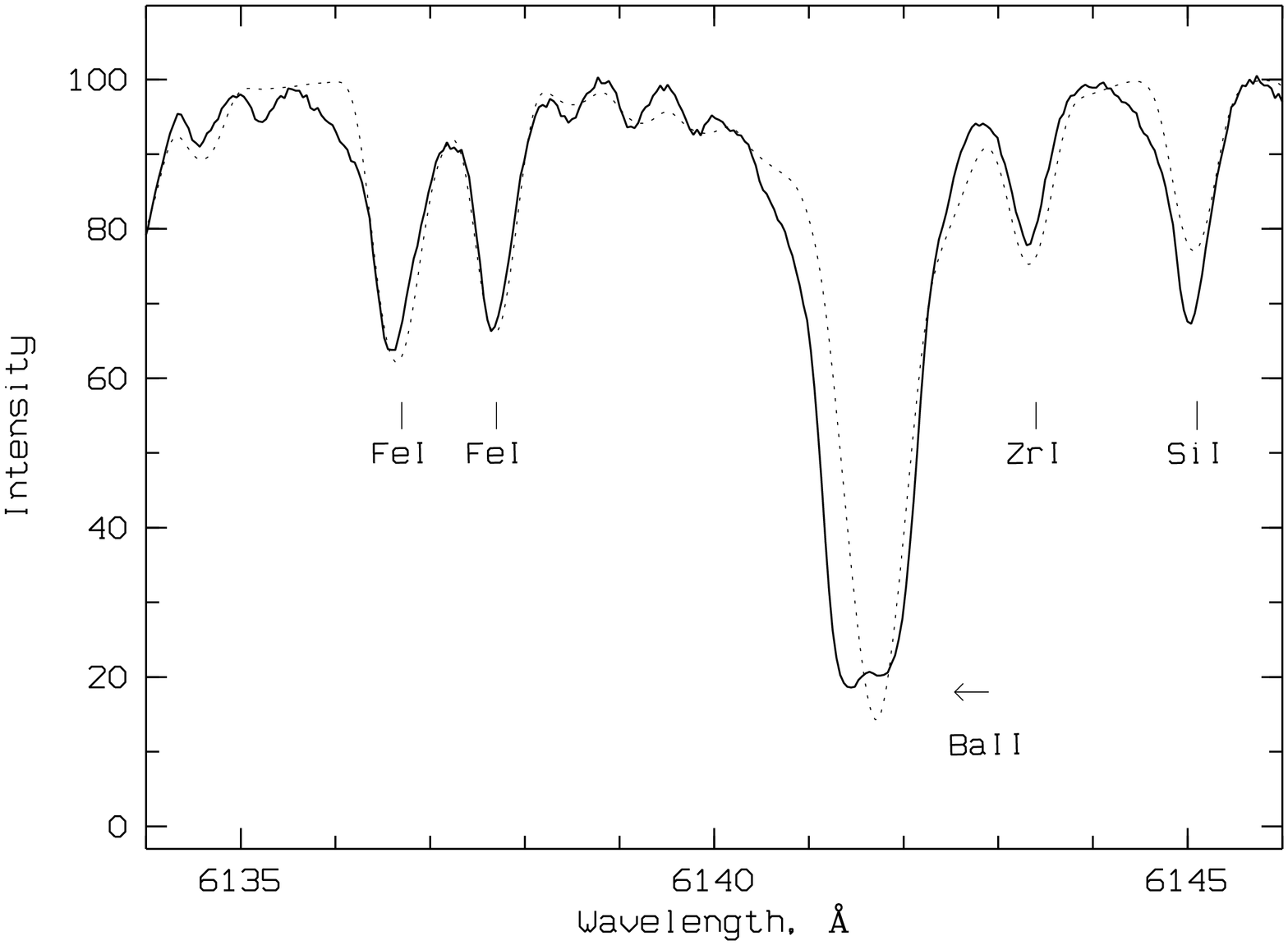,angle=-90,width=15cm,height=10cm}
\caption{Part of the spectrum for V354\,Lac. The Fe\,I\,6136.7,
         Fe\,I\,6137.7, Ba\,II\,6141.7, Zr\,I\,6143.4, and Si\,I\,6145.1\,\AA{}
	 lines are marked. The dotted line indicates the theoretical spectrum calculated
          with T${\rm eff}=5650$\,K, $\log g = 0.2$, $\xi_{\rm t}=5.0$\,km$/$s,
	  and the elemental abundances derived previously (Klochkova {\it et~al.} 2009).}
\end{figure}

Hrivnak (1995) classified V354\,Lac as a G5\,Iap supergiant. The main
features of its optical spectrum were pointed out even in the firrst works
with a low spectral resolution. Hrivnak (1995) and Hrivnak \& Kwok (1991)
found that, compared to the spectrum of a normal supergiant with a similar
temperature, the spectrum of V354\,Lac exhibits a weaker H$\delta$ line,
stronger Ba\,II lines, and CN, C$_2$ and C$_3$ molecular absorption
bands. Most of the listed features are also observed in our 2007--2008
spectra. In particular, we used rotational lines of the Swan C$_2$\,(0;0)
band, with the head at 5165.2\,\AA{}, to determine the expansion velocity
of the envelope (for more detail, see below).

The low-excitation Ba\,II lines are the strongest absorption features in
the spectrum of V354\,Lac; their equivalent widths W$_{\lambda}$ exceed
0.6\,\AA{}. The absorption features of other ions of s-process elements
(La, Ce) are equally strong; their W$_{\lambda}>$0.3\,\AA. Figure\,1 shows
the H$\alpha$ line profile that consists of an absorption component with a
narrow core and broad wings for JD\,=\,2454225.5. As we see from the
figure, the observed H$\alpha$ line profile in the spectrum of V354\,Lac
agrees with the theoretical one calculated with its fundamental parameters
T$_{\rm eff}$\,=\,5650 K, $\log g$\,=\,0.2, $\xi_t$\,=\,5.0\,km$/$s, and
the elemental abundances derived by Klochkova {\it et~al.} (2009). Thus, we
found no weakening of this line, as expected from the results of Hrivnak
(1995). This is indicative of the line formation in the stellar
photosphere and a weak contribution from the envelope. The positions of
the H$\alpha$ and H$\delta$ cores differ by 2--4\,km$/$s from the
averaged velocity measured from photospheric metal lines.

The high spectral resolution allowed us to detect another, previously
unobservable feature of the optical spectrum for V354\,Lac --- splitting
of the cores of the strongest heavy-metal lines. This splitting is clearly
seen from Fig.\,2 for the profile of the Ba\,II\,$\lambda$\,6141\,\AA{}
line with an equivalent width W$_{\lambda}\approx$\,1\,\AA{}. Such
splitting (or asymmetry of the line profile due to the extended blue wing)
is also observed for other Ba\,II lines ($\lambda$~5435, 5853, and
6496\,\AA{}) and for such strong lines as YII\,5402\,\AA{},
La\,II\,6390\,\AA{}, and Nd\,II\,5234\,\AA{} and 5293\,\AA{}. The lines of
these heavy elements in the spectrum of V354\,Lac are enhanced to an
extent that their intensities are comparable to those of HI lines (compare
Fig.\,1 and 2). Asymmetry is clearly seen, for example, in the
Ba\,II\,5853 line profile (Fig.\,3). This figure, which shows the
Ba\,II\,5853 and 6141\,\AA{} lines for several dates of observations, also
illustrates variability of the profiles of strong lines. Unfortunately,
the Ba\,II\,6141\,\AA{} line was recorded only for two dates, but the
profiles may be said to be variable even in this case.

All of the lines with detected core splitting (or profile asymmetry) are
distinguished by low lower-level excitation potentials, $\chi_{\rm low}
<$\,1\,eV. Obviously, the strong low-excitation lines originating in the
upper layers of the stellar atmosphere are affected by the gas envelope.
As an illustration, Fig.\,4 compares the Ba\,II and La\,II lines in the
spectrum taken on one date (JD\,=\,2454225.51).

As an example, let us consider the picture of core splitting for
heavy-element lines in more detail for the Ba\,II\,6141\,\AA{} line, for
which this effect is most pronounced. The separation between the absorption
components of the Ba\,II\,6141\,\AA{} line is about 35\,km$/$s. The
short-wavelength component coincides in position with the circumstellar
component of the NaD1 profile (see Fig.\,5). This coincidence confirms 
that, apart from the photospheric component, the complex  
Ba\,II\,6141\,\AA{} line profile contains a component originating in the
circumstellar envelope. At an insuficient spectral resolution, the
intensity of the envelope components is added to the intensity of the
components originating in the atmosphere. As follows from Zacs {\it
et~al.} (1995), Reddy {\it et~al.} (2002), and Klochkova {\it et~al.}
(2009), large overabundances of heavy elements synthesized during the
s-process are observed in the atmosphere of V354\,Lac. Because of core
splitting (and/or asymmetry), the heavy-element abundances derived from
strong absorption lines in the spectrum of V354\,Lac turn out to be
overestimated by about 0.2--0.4\,dex at an unsufficient spectral resolution.

\begin{figure}[h]
\psfig{figure=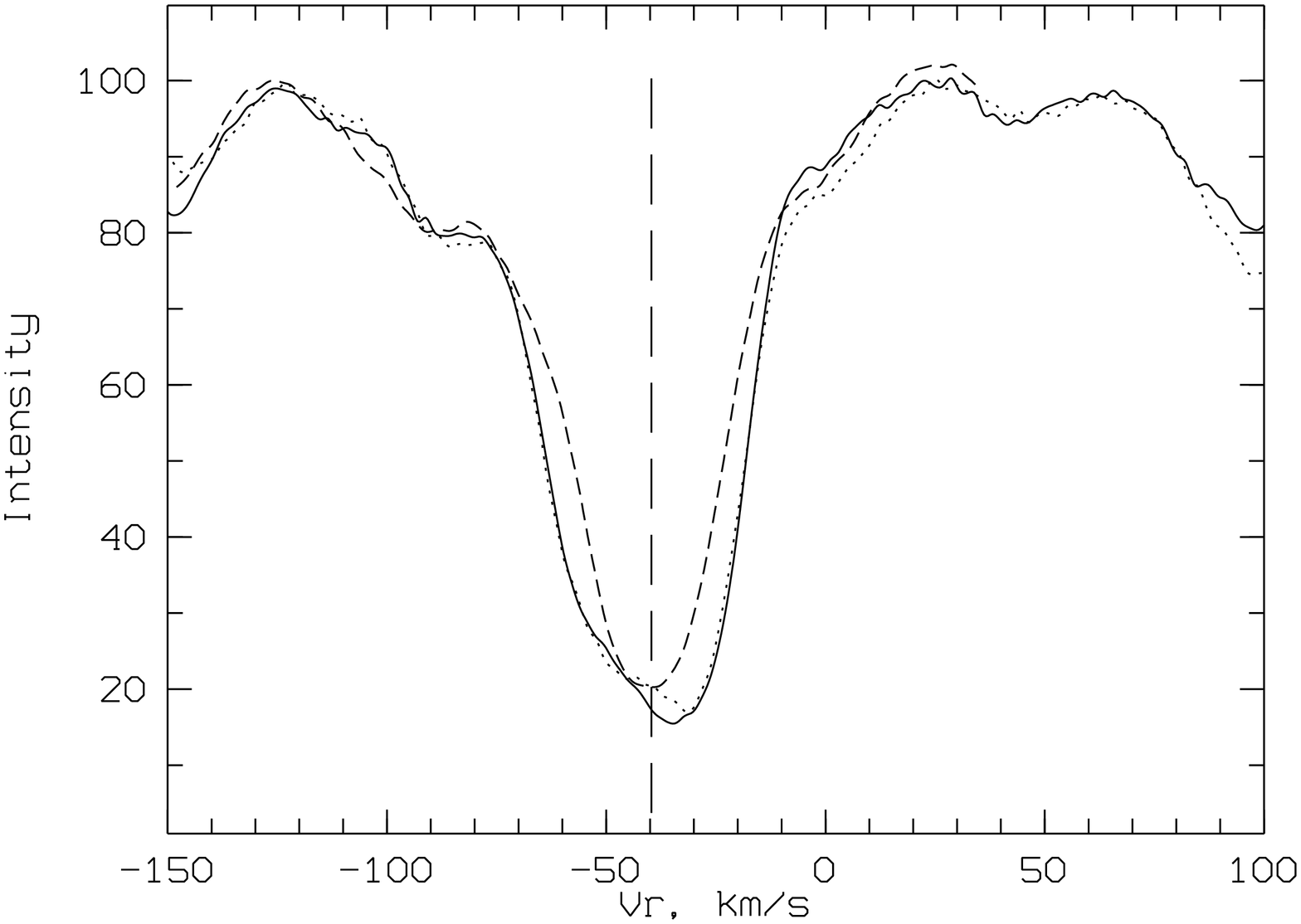,angle=-90,width=14cm,height=9cm}
\psfig{figure=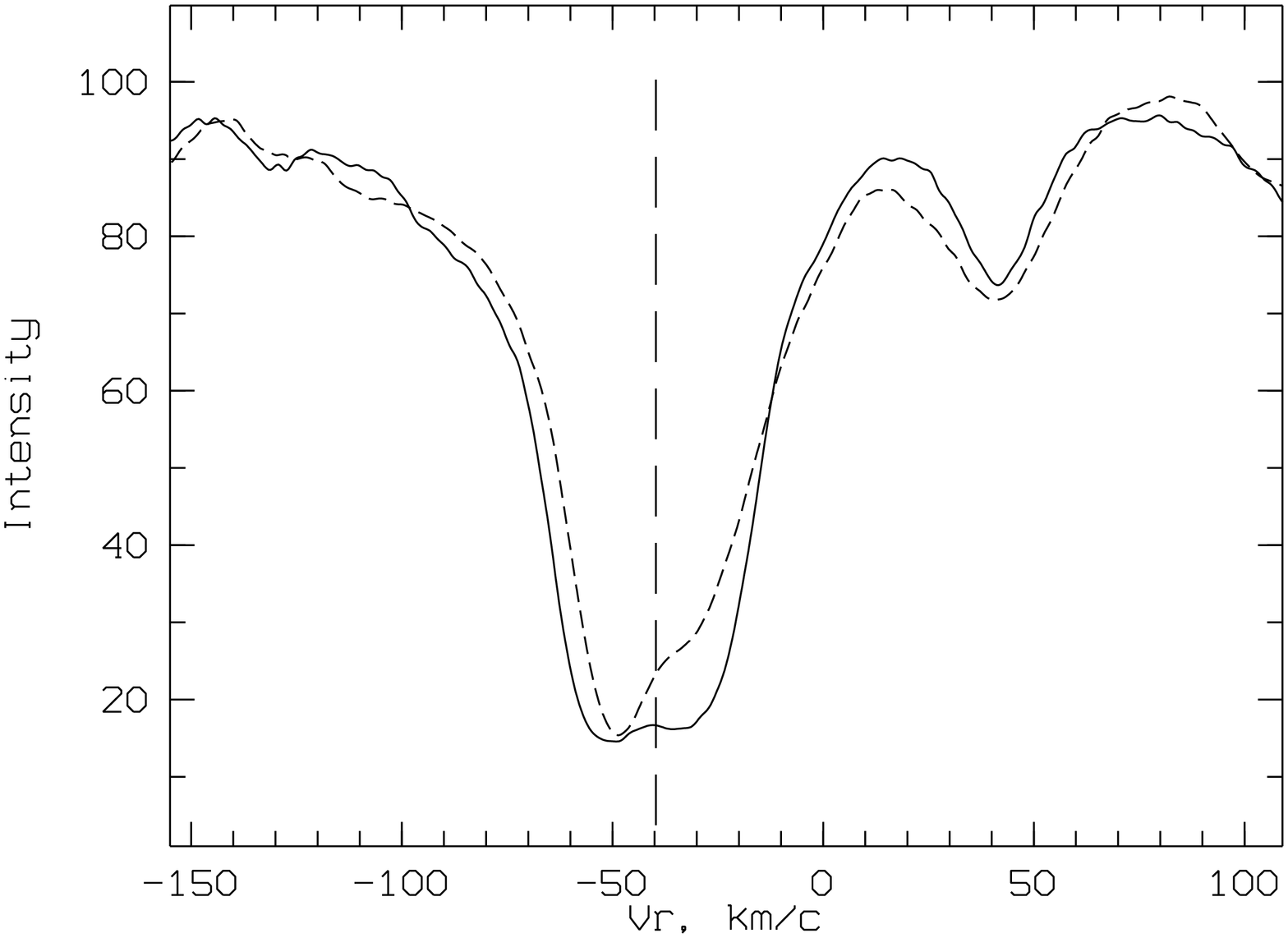,angle=-90,width=14cm,height=9cm}
\caption{Variability of the Ba\,II\,5853\,(a) and 6141\,\AA{}\,(b)
         line profiles in the spectra of V354\,Lac: the dotted, solid,
	 and dashed lines are for JD\,=\,2454170.6, JD\,=\,2454225.5,
	 and JD\,=\,2454727.4, respectively. The systemic velocity is indicated
	 by the vertical dashed line.}
\end{figure}

\begin{figure}[hbtp]
\psfig{figure=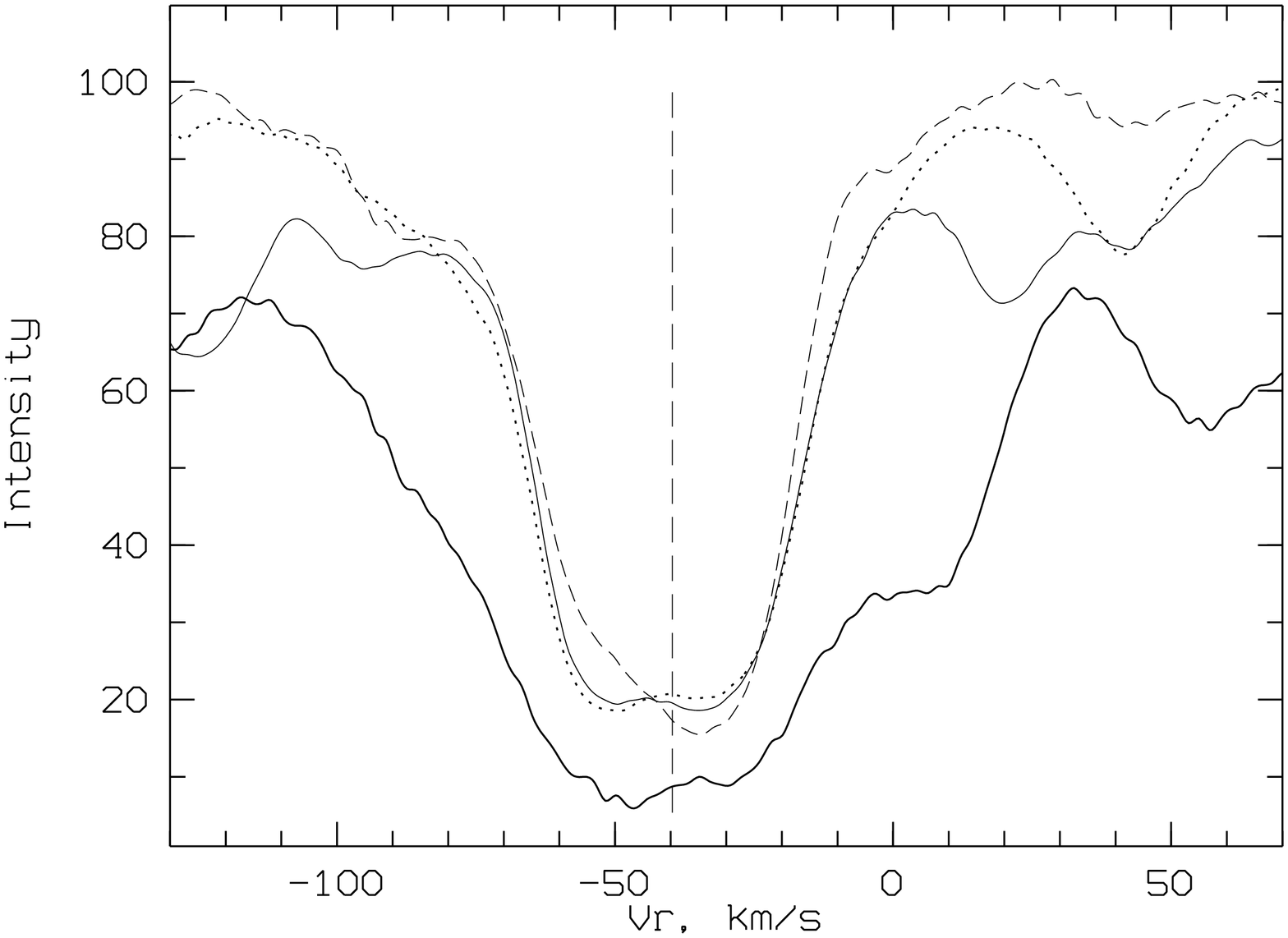,angle=-90,width=14cm,height=9cm}
\psfig{figure=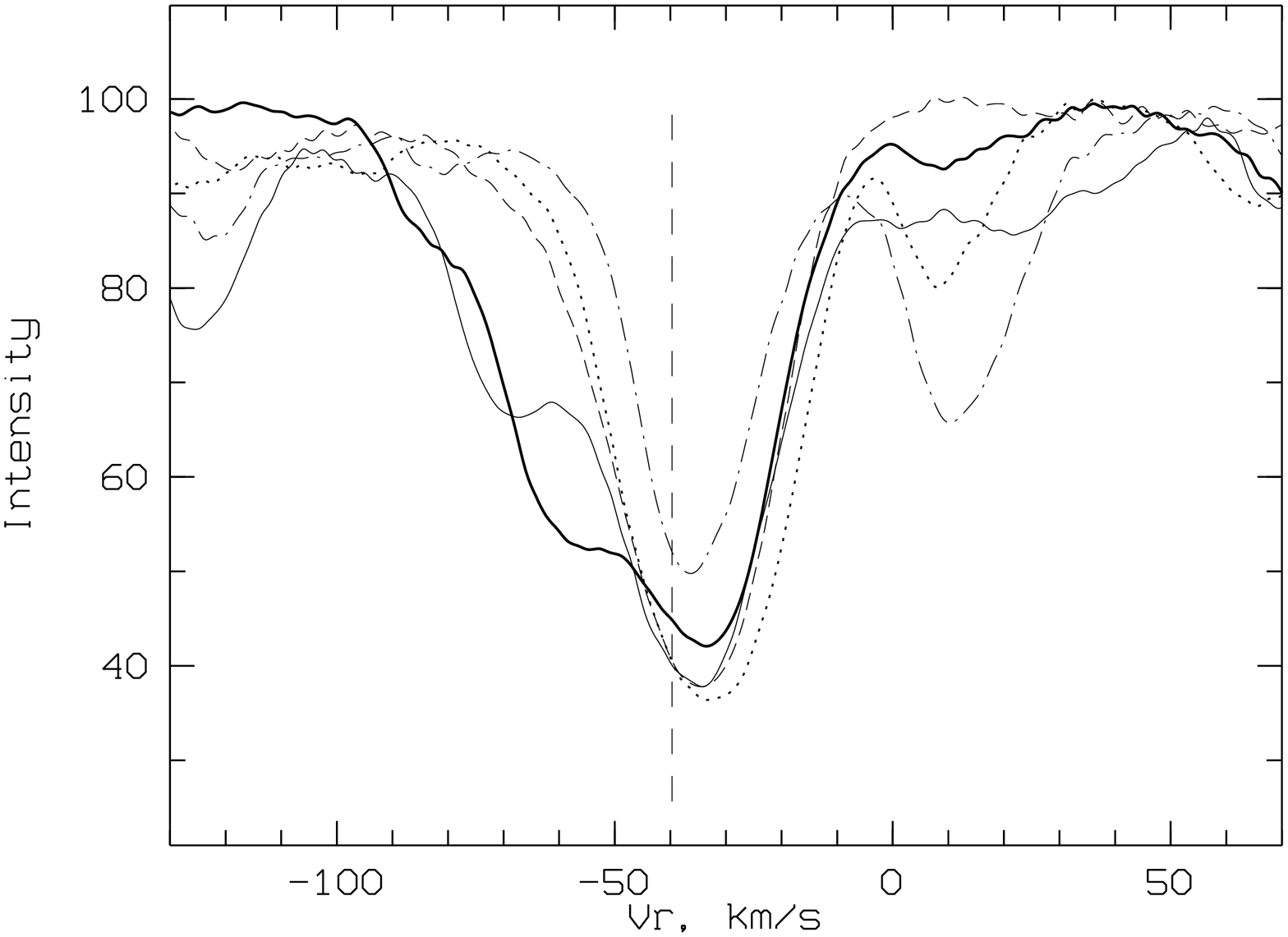,angle=-90,width=14cm,height=9cm}
\caption{(a) Ba\,II line profiles in the spectrum of V354\,Lac on
        JD\,=\,2454225.5: the thick solid, thin solid, dashed, and dotted
	lines represent Ba\,II\,4934\,\AA{}, Ba\,II\,6496\,\AA{},
	Ba\,II\,5853\,\AA{}, and Ba\,II\,6141\,\AA{}, respectively.
	(b) The same as in panel (a) but for the La\,II\,lines:
	the thick solid, thin solid, dash--dotted, dashed, and dotted lines
	represent La\,II\,6390\,\AA{}, La\,II\,6320\,\AA{}, La\,II\,5808\,\AA{},
	and La\,II\,6526\,\AA{} and La\,6262\,\AA{}, respectively.
	The systemic velocity is indicated by the vertical dashed line. }
\end{figure}

\begin{figure}[hbtp]
\psfig{figure=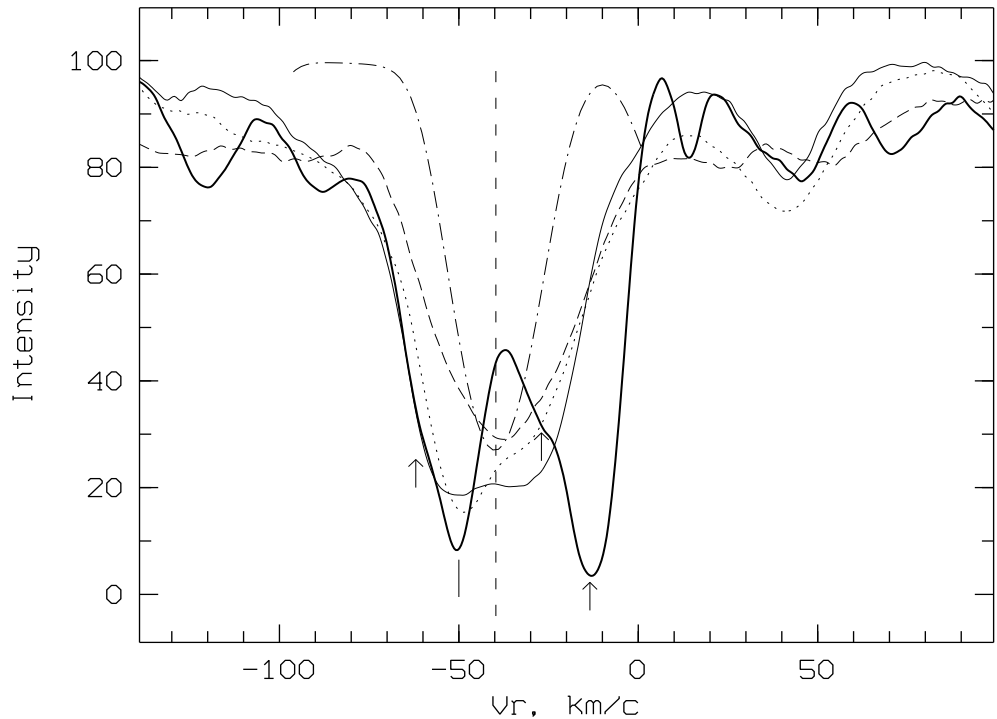,angle=-90,width=15cm,height=10cm}
\caption{Selected line profiles in the spectrum of V354\,Lac: the thick
        curve represents the NaD1 line of the sodium doublet; the thin and
        dotted curves represent the Ba\,II\,6141\,\AA{} line for two dates; the
        dashed curve represents H$\alpha$. The dash--dotted line indicates the
        theoretical NaD1 profile calculated with T$_{\rm eff}$\,=\,5650\,K,
	$\log g = 0.2$, $\xi_{\rm t} = 5.0$\,km$/$s, and the elemental
	abundances from Klochkova {\it et~al.} (2009). The vertical dashed line
	indicates the systemic velocity; the arrows mark the interstellar
	components; the vertical bar marks the circumstellar component of the NaD1 line.}
\end{figure}

The abundances derived from moderate-intensity lines will be more
realistic. We emphasize that such a complex profile, which, apart from
the photospheric and interstellar components, also contains the
circumstellar component, has been previously observed only for the NaI\,D
doublet lines. In particular, this is also true for the star V354\,Lac
under study. Reddy {\it et~al.}~(2002) distinguished the circumstellar component
in the NaI\,D lines. The circumstellar absorption features of the NaI\,D
lines were also identified in the spectrum of the well-studied post-AGB
star HD\,56126 (Bakker {\it et~al.} 1996; Klochkova \&~Chentsov 2007). In
addition, manifestations of the gas--dust circumstellar envelope in the
form of emission components in the NaI\,D lines are known. As an example,
we present the spectra of the protoplanetary nebula V510\,Pup identified
with the infrared source IRAS\,08005$-$2356 (Klochkova \&~Chentsov, 2004),
the two-lobe nebula Egg\,=\,AFGL\,2688 (Klochkova {\it et~al.} 2004b), and
the semiregular variable QY\,Sge\,=\,IRAS\,20056+1834 (Rao {\it et~al.} 2002;
Klochkova {\it et~al.} 2007c). In our study, we have found a manifestation of
the envelope in heavy-metal lines for the first time.

\subsection{Radial velocities}

{\bf Metal lines.} To determine the mean heliocentic radial velocity
V$_\odot$, we measured the positions of a large number (about 300) of
minimally blended absorption lines in the spectra of V354\,Lac. The lines
were selected using a spectral atlas for the post-AGB star HD\,56126,
which may be considered as a canonical post-AGB object (Klochkova {\it
et~al.} 2007a). The atlas was compiled by Klochkova {\it et~al.} (2007b)
from echelle spectra taken with the same NES spectrograph of the 6-m
telescope. The blending level in the spectrum of V354\,Lac is higher than
that in the spectrum of HD\,56126 because of the later spectral type of
the star (the effective temperature of HD\,56126, T$_{\rm eff}= 7000$\,K,
was determined by Klochkova (1995)) and because of the splitting and
asymmetry of many lines. Due to enhanced blending, the accuracy of
measuring V$_{\odot}$ using one line (rms deviation $\sigma$) is about 2
km$/$s from our 2007--2008 spectra. The radial velocities of V354\,Lac
measured from the set of spectral features are listed in the table. Here,
the second column gives the mean V measured from metal lines. The next
columns give V$_{\odot}$ measured from the H$\alpha$ and H$\beta$ lines,
the components of the NaD lines, and the rotational lines of the Swan
C$_2$ molecular band.

In addition to the data on the 2007--2008 spectra, the table includes our
measurements of V$_{\odot}$ based on a spectrum taken previously at BTA
with the Lynx echelle spectrograph (Panchuk {\it et~al.} 1999) with
resolution R\,=\,25000. The last row gives the mean V$_{\odot}$ from Reddy
{\it et~al.} (2002). As follows from the longterm observations by Hrivnak
and Lu (2000), the radial-velocity variability amplitude and period for
V354\,Lac are typical of PPN: the radial velocity varies within the range
($-34 \div -41$)\,km$/$s with a period of 127$^{\rm d}$. All of the mean
V$_{\odot}$ derived using metal absorption lines from the table lie within
this range of variability.

\vspace{6mm}
\begin{table}[t]
\caption{Heliocentric radial velocities V$_{\odot}$  for three epochs of
         observations measured from various spectral features. The number
	 of measured lines is given in parentheses. The first row provides
	 V$_{\odot}$  derived from the spectrum taken at BTA with the Lynx
	 spectrograph (Panchuk {\it et~al.} 1999, R\,=\,25000).
         The last row presents the data from Reddy {\it et~al.} (2002)}
\begin{tabular}{r|l|c|l|c|l}
\hline
JD=24...    &\multicolumn{5}{c}{V$_{\odot}$, km$/$s} \\
\cline{2-6}
            & metals & HI &\multicolumn{2}{c|}{NaD} & C$_2$    \\ [3pt]
\cline{4-5}
            & & & blue & red &      \\
\hline
48850.51    &$-$38.2  &$-$41.1 H$\alpha$&$-$50.2 &$-$13.2&$-$50.5(8)   \\ [3pt]
54170.58    &$-$40.1  &$-$45.1 H$\beta$ &$-$50.6 &$-$14.4&$-$50.1(21)   \\ [3pt]
54225.51    &$-$38.4  &$-$37.2 H$\alpha$&$-$51.1 &$-$13.4&            \\ [3pt]
54727.35    &$-$38.0  &$-$34.6 H$\alpha$&$-$51.6 &$-$14.0&            \\ [3pt]                           
\hline 
20.08.2000  &\multicolumn{4}{l}{$-$42.4 (Reddy {\it et~al.} 2002)} \\  [3pt]
\hline
\end{tabular}
\end{table}

{\bf Molecular spectrum.} Apart from the infrared excess and reddening,
the presence of a gas--envelope around the central star of PPN also
manifests itself in features of the optical spectra. Since the molecular
bands can be formed in the atmosphere of a star with a temperature T$_{\rm
eff}<3000$\,K, it is obvious that for a G5 star, the molecular bands are
formed in the circumstellar envelope. Vibrational Swan C$_2$ molecular
bands are observed in our spectra of V354\,Lac. The high spectral
resolution makes it possible to measure accurately the positions of the
rotational lines of the Swan~(0;0) band. Using the rotational-line
wavelengths from the electronic tables to the paper by Bakker {\it et~al.}
(1997), we measured the positions of 21 rotational lines of the Swan~(0;0)
band and determined the mean radial velocity in the band-formation region,
V$_{\odot}(0;0)=-50.1 \pm 0.2$\,km$/$s. Because of their narrow profiles
compared to the photospheric lines, the rotational lines of the Swan~(0;0)
band are easily distinguished in the spectrum. Therefore, the position of
one line can be measured with an accuracy of about 0.8\,km$/$s, which is
much better than that for photospheric absorption lines. Since the
Swan~(1;0) band in the short-wavelength part of the spectrum
4712--4734\,\AA{} is strongly blended by photospheric lines, the
measurement accuracy is much lower, V$_{\odot}(1;0)=-50.0\pm 1.0$\,km$/$s.

The shift of circumstellar features relative to the systemic velocity
allows the expansion velocity of the corresponding envelope regions to be
determined. Fong {\it et~al.} (2006) derived the systemic velocity V$_{\rm
lsr}^{\rm sys}=-27.5$km$/$s (V$_{\odot}^{\rm lsr} =-39.7$\,km$/$s) of the
source IRAS\,22272+5435 from the position of the center of the CO~(1--0)
profile. In contrast to the emission CO lines formed in an extended
envelope that expands in all directions, the observed absorption lines of
molecular carbon are formed in the part of the envelope located between
the star and the observer. As a result, we obtain the velocity of the
Swan-bands formation region relative to the stellar center, V$_{\rm
exp}$\,=\,10.8\,km$/$s. This value, which may be considered as the
envelope expansion velocity derived from optical spectra, agrees well with
the the expansion velocity for IRAS\,22272+5435, V$_{\rm exp}=10.8 \pm
1.1$\,km$/$s, from the catalog by Loup {\it et~al.} (1993), who collected
numerous observations of circumstellar envelopes in CO and HCN molecular
bands. Note that the envelope-expansion velocity for IRAS\,22272+5435 is
typical of the circumstellar envelopes of post-AGB stars (see, e.g., Loup
{\it et~al.} 1993). 

The heliocentric systemic velocity V$_{\odot}^{\rm lsr} =-39.7$\,km$/$s
agrees well with the mean velocity of the star inferred from metal lines.
This agreement indicates that there is no secondary component in the
system of IRAS\,22272+5435 or, to be more precise, there is no secondary
component with a stellar mass. This is a nontrivial result, since the
chemical evolution, mixing, and dredge-up of nuclear-reaction products to
the surface layers of the stellar atmosphere, the outflow of matter, and
the formation of envelope morphology can proceed in a special way in the
presence of a secondary companion.

Taking into account the galactic CO velocity maps (Vall\`ee 2008), the
galactic coordinates (l\,=\,103$\lefteqn{.}^{\rm o}$3,
b=\,$-2\lefteqn{.}^{\rm o}$51), and the systemic velocity V$_{\rm
lsr}^{\rm sys}=-27.5$km$/$s of IRAS\,22272+5435, we can assume that the
source is located between the local and Perseus arms.

{\bf NaD doublet.} Both lines of the resonance NaI doublet in the spectrum
of V354\,Lac have a complex structure. As follows from the table and
Fig.\,5, which shows the D1 line profile, the doublet lines contain two
strong absorption components whose positions correspond to the velocities
V$_{\odot}$\,=\,$-50$ and $-13$\,km$/$s. Obviously, the line with
V$_{\odot}$\,=\,$-50$\,km$/$s originates in the circumstellar envelope, where
the circumstellar Swan C$_2$ molecular bands are also formed.
The second component with (V$_{\odot}$\,=\,$-13$\,km$/$s) is interstellar
in origin. The presence of this interstellar component, V$_{\rm lsr}
\approx-27$\,km$/$s, confirms our assumption that V354 Lac is located
in the Galaxy farther than the local arm.
According Georgelin and Georgelin (1970), the radial velocity in the local
and Perseus spiral arms of the Galaxy are V$_{\rm lsr}\approx -10$\,km$/$s
and $-55$\,km$/$s, respectively. Thus, the distance to the Perseus arm
d\,=\,3.6\,kpc derived by Foster \&~MacWilliams (2006) may be used as
an upper limit for the source. By modeling the bolometric flux from
IRAS\,22272+5435, Loup {\it et~al.} (1993) estimated the distance to the source
to be d\,=\,2.35\,kpc.

As follows from Fig.\,5, the blue wings of the  NaI absorption
features hint at the presence of poorly resolvable components in our
spectra: V$_{\odot}\approx -57$\,km$/$s (V$_{\rm lsr}\approx -70$\,km$/$s)
and V$_{\odot}\approx -24$\,km$/$s (V$_{\rm lsr}\approx -37$\,km$/$s).

\section*{Conclusions}

Based on the optical spectra of the post-AGB star V354 Lac
taken in 2007--2008 with the echelle spectrograph of the 6-m telescope
with spectral resolution R\,=\,60000, we detected core splitting or
asymmetry (an extended blue wing) of absorption lines with lower-level
excitation potentials $\chi_{\rm low}<$1\,eV. This primarily applies to the
strongest absorption lines identified with heavy-metal (Ba, La, Ce, Nd)
ion lines. Allowance for the detected core splitting of the strongest
absorption lines reduces the heavy-metal overabundances revealed
previously by 0.2--0.4 dex.

The observed H$\alpha$ line profile is in good agreement with the
theoretical one calculated with the fundamental stellar parameters. This
is indicative of the line formation in the stellar photosphere and a weak
contribution from the envelope. The radial velocity of the star measured
for two epochs of observations in 2007--2008 closely coincides, within the
error limits, with the previously published data. This suggests that there
are no velocity field variations in the atmosphere and circumstellar
envelope of V354\,Lac over the last 15 years of observations.

\subsection*{Acknowledgments}

This work was supported by the Russian Foundation for Basic Research
(project No.\,08--02--00072\,a), the ``Extended objects in the Universe''
Basic Research Program of the Division of Physical Sciences of the Russian
Academy of Sciences, and the ``Origin and Evolution of Stars and
Galaxies'' Program of the Presidium of the Russian Academy of Sciences.

\newpage

\centerline{\Large\bf References}

\begin{enumerate}

\item{} M. Asplund, N. Grevesse, \& A.J. Sauval, ASP Conf. Ser.
           \textbf{336}, 25 (2005).

\item{}  E. J. Bakker, E. F. van Dishoeck, L. B. F. M. Waters, \&
      T.~Schoenmaker, Astron. Astrophys. \textbf{323}, 469 (1997).

\item{} E. J. Bakker, L. B. F. M. Waters, H. J. G. M. Lamers, {\it et~al.},
        Astron. \&~Astrophys. \textbf{310}, 893 (1996).

\item{}  T. Blocker, Astrophys. \& Space Sci. \textbf{275}, 1 (2001).

\item{}  D. Fong, M. Meixner, E. C. Sutton, {\it et~al.}, Astrophys. J. \textbf{652},
        1626 (2006).

\item{}  T. Foster \& J. MacWilliams, Astrophys. J. \textbf{644}, 214 (2006).

\item{}  G. A. Galazutdinov, Preprint SAO No.\,92, (1992).

\item{}  Y. P. Georgelin \&~Y. M. Georgelin, Astron. \& Astrophys.
          \textbf{6}, 349 (1970).

\item{}  B. J. Hrivnak, Astrophys. J. \textbf{438}, 341 (1995).

\item{} B. J. Hrivnak \&~S. Kwok, Astrophys. J. \textbf{371}, 631 (1991).

\item{} B. J. Hrivnak \&  Wenxian~Lu, IAU Symp. No.\,177, Ed.~by R.~F.~Wing
     (Kluwer Acad., Dordrecht, 2000), p. 293.

\item{}  C. S. Jeffery \& D. Schonberner, Astron. \&~Astrophys.
         \textbf{459}, 885 (2006).

\item{} V. G. Klochkova, Mon. Not. R. Astron. Soc. \textbf{272}, 710 (1995).

\item{} V. G. Klochkova \& E. L. Chentsov.  Astron. Rep. 48,
         \textbf{301}, (2004).

\item{} V. G. Klochkova \& E. L. Chentsov,  Astron. Rep.
             \textbf{51}, 994 (2007).

\item{} V. G. Klochkova, M. V. Yushkin, A. S. Miroshnichenko, {\it et~al.},
            Astron. \&~Astrophys. \textbf{392}, 143 (2002).

\item{} V. G. Klochkova, V. E. Panchuk, \&~N. S. Tavolzhanskaya,
             Astron. Lett. \textbf{28}, 49 (2002).

\item{} V. G. Klochkova, E. L. Chentsov, V. E. Panchuk, \&~M.~V.~Yushkin,
    Inform. Bull. Var. Stars 5584, 1 (2004a).

\item{} V. G. Klochkova, V. E. Panchuk, M. V. Yushkin, \&~A.~S.~Miroshnichenko,
          Astron. Rep. \textbf{48}, 288 (2004b).
 
\item{} V. G. Klochkova, E. L. Chentsov, V. E. Panchuk, {\it et~al.}, Baltic Astron.
            16, \textbf{155} (2007a).

\item{} V. G. Klochkova, E. L. Chentsov, N. S. Tavolganskaya, \& M.~V.~Shapovalov,
     Bull. Spec. Astrophys. Observ. \textbf{62}, 162 (2007b).

\item{} V. G. Klochkova, V. E. Panchuk, E. L. Chentsov, \& M.~V.~Yushkin,
 Bull. Spec. Astrophys. Observ. \textbf{62}, 233 (2007c).

\item{} V. G. Klochkova, E. L. Chentsov, \& V. E. Panchuk, Bull.
     Spec. Astrophys. Observ. \textbf{63}, 112 (2008).

\item{} V. G. Klochkova, V. E. Panchuk, \& N. S. Tavolganskaya,
       Bull. Spec. Astrophys. Observ. \textbf{64}, 155 (2009).

\item{} C. Loup, T. Forveille, A. Omont, \& J. F. Paul, Astron. \&~Astrophys.
        Suppl. Ser. 99, \textbf{291} (1993).

\item{} V. E. Panchuk, V. G. Klochkova, I. D. Naidenov, {\it et~al.}, Preprint SAO
      No.\,139 (1999).

\item{} V. Panchuk, V. Klochkova, M. Yushkin, \& I.~D.~Najdenov,
     In: ``The UV Universe: Stars from Birth to Death'', Proc. of the Joint
      Discussion No. 4 during the IAU General Assembly of 2006, Ed. by
      A.~I.~Gomez de Castro \& M.~A.~Barstow (2007), p. 179.

\item{} N. Kameswara Rao, A. Coswami, \& D.~L.~Lambert, Mon. Not. R. Astron.
       Soc. 334, \textbf{129} (2002).

\item{} B. E. Reddy, D. Lambert, G. Gonzalez, \& D. Yong, Astrophys. J.
         \textbf{564}, 482 (2002).

\item{} J. P. Vall\`ee, Astron. \textbf{J}. 135, 1301 (2008).

\item{} M. V. Yushkin \&~V. G. Klochkova, Preprint SAO No.\,206 (2005).

\item{} L. Zacs, V. G. Klochkova, \& V. E. Panchuk, Mon. Not. R.
               Astron. Soc. \textbf{275}, 764 (1995).

\end{enumerate}

\end{document}